\documentclass[aps,tightenlines,showpacs]{revtex4}
\usepackage[dvips]{graphicx}
\usepackage{latexsym}
\usepackage{amsmath}

\newcommand{\BE}{\begin{equation}}
\newcommand{\EE}{\end{equation}}
\newcommand{\BA}{\begin{eqnarray}}
\newcommand{\EA}{\end{eqnarray}}

\begin{document}

\title{Gaussian effective potential for the standard model 
\\SU(2)$\times$U(1) electroweak theory}

\author{Fabio Siringo}
\author{Luca  Marotta}
\affiliation{Dipartimento di Fisica e Astronomia, 
Universit\`a di Catania,\\
INFN Sezione di Catania and CNISM Sezione di Catania,\\
Via S.Sofia 64, I-95123 Catania, Italy}

\date{\today}
\begin{abstract}
The Gaussian Effective Potential (GEP) is derived for the non-Abelian
SU(2)$\times$U(1) gauge theory of electroweak interactions. First
the problem of gauge invariance is addressed in the Abelian U(1)
theory, where an optimized GEP is shown to be gauge invariant. The method
is then extended to the full non-Abelian gauge theory where, at variance with
naive derivations, the GEP is proven to be a genuine variational tool in
any gauge. The role of ghosts is discussed and the unitarity gauge
is shown to be the only choice which allows calculability without insertion
of further approximations. The GEP for the standard model is derived and
its predictions are compared to the known phenomenology, thus showing that
the GEP provides an alternative non-perturbative description 
of the known experimental data. By a consistent renormalization of masses the
full non-Abelian calculation confirms
the existence of a light Higgs boson in the non-perturbative strong coupling regime
of the Higgs sector.
\end{abstract}


\maketitle

\section{introduction}

It is now widely believed that
the Higgs sector of electroweak interactions can be described
by a scalar field with a self-interaction which could be large enough to raise
some doubt on the validity of standard perturbative approaches. 
Thus, while perturbative
results might be questioned, non-perturbative calculations would be required at least
for comparison.
Variational calculations are usually quite reliable for describing
strong coupling phenomena, but their use in quantum field theory must face
several difficult problems\cite{Feynman}.
The problem of calculability can only be solved by use of a Gaussian wave functional,
which has its merits as discussed in several papers on the Gaussian Effective Potential
(GEP)\cite{schiff,rosen,barnes,kuti,chang,weinstein,huang,bardeen,peskin,stevenson}.
An other important problem is the predominance of high momentum fluctuations in the
vacuum espectation values. However the standard model of electroweak interactions
is usually regarded as an effective model with a finite energy cut-off which
regulates the theory. Thus the role of high momentum fluctuations is in part
reduced, as the predictions of the GEP on effective models have been found to
be reliable when compared to experimental results\cite{marotta,camarda,siringo_sigma}.

It has been pointed out that gauge invariance could be an other important challange for
variational calculations, as there is no way to build a gauge invariant Gaussian functional
in non-Abelian gauge theories\cite{kogan95}. It has been argued that in principle, 
if the states are not gauge invariant, 
they could be unphysical and span a larger Hilbert space where the unphysical
energies could even be lower than the true physical vacuum\cite{kogan95}.
However in this paper we show that a genuine variational GEP can be found for the
non-Abelian SU(2)$\times$U(1) standard model of electroweak interactions, and that
for any chosen gauge the GEP can be proven to stay above the true effective potential. 
The genuine variational nature of the GEP makes the choice of gauge a question of
taste and numerical convenience, and the physical unitarity gauge may be used without
affecting the variational nature of the calculation.

Some further motivation for the work arises from a successful attempt 
to explain mass generation
in the minimal left-right symmetric model of 
electroweak interactions\cite{siringo06,siringo04},
where two scalar Higgs doublets and no bidoublet are present. At tree level
that model predicts a vanishing expectation value for one of the scalar Higgs doublets,
and that is a problem since all the fermionic masses turn out to be 
vanishing as well\cite{siringoPRL}.
In that framework quantum fluctuations have been studied by the GEP and 
shown\cite{siringo06}
to destabilize the symmetric vacuum towards a physical finite expectation value for
both the Higgs doublets. While those findings are compatible 
with the phenomenology, their accuracy could be questioned for the neglection of
all the weak couplings. Actually it was a simplified Abelian 
toy model, with only Higgs and fermionic fields.
Thus an extension of the GEP method to the full non-Abelian SU(2)$\times$U(1) gauge
group would allow for quantitative predictions in the standard model and in
its minimal left-right symmetric versions.

We must mention that this is not the first attempt to apply the GEP to the 
non-Abelian gauge theory, as previous naive calculations have been reported.
It is very important to stress that the reliability of a variational calculation
requires that no uncontrolled approximation should be added. The main result
of this paper is the rigorous proof of the genuine variational nature of
the GEP in unitarity gauge. In order to avoid problems regarding the gauge dependence
of the Hamiltonian, we derive the GEP in the Lagrangian formalism and start
from a fully gauge invariant vacuum to vacuum transition amplitude. As in previous
works on
the U(1) theory\cite{ibanez,camarda,marotta}, the GEP is derived 
by a systematic use of Jensen's inequality for expectation values of convex functions.
As a consequence the GEP can never fall below the exact effective potential, and
its minimum yields the best approximation to the vacuum energy density.

The derivation is useful for clarifying the role played by any gauge choice. In fact
Jensen's inequality does not hold for Grassmann anticommuting fields and when
ghost fields are present the naive use of the GEP turns out to be a tree level
perturbative approximation. Thus the gauge must be properly chosen in order to
avoid the presence of ghosts, and unitarity gauge turns out to be a good choice.
Moreover we show that, even for the U(1) group, Jensen's inequality works at
its best for some special gauge choice. In the non-Abelian theory the best gauge
turns out to be the Unitarity one which also yields a very clear physical
description.

In the standard model of weak interactions the method is shown to be a useful
non-perturbative tool for the study of the Higgs sector in the strong coupling regime.
The GEP provides a consistent renormalization of masses, and predicts the possible
existence of a light Higgs boson even if the self-coupling were very large.
In other words a light Higgs boson would not rule out a very large self-coupling which
would question most of the perturbative calculations. The role of gauge interactions on
the Higgs sector is also discussed and shown to be very small, as expected.

The paper is organized as follows: 
in section II the use of Jensen's inequality is discussed for the
Abelian U(1) theory, and the resulting optimized GEP is shown to
be gauge invariant; in section III the full non-Abelian SU(2)$\times$U(1)
gauge group is considered, and the main lines of the GEP derivation are outlined;
in section IV the GEP is derived for the standard model of electroweak interactions,
and in section V the gap equations are discussed in detail; section VI deals with
a simple modified variational approach which allows a direct comparison with the
phenomenology, while a consistent renormalization of masses is addressed in section
VII where the phenomenological predictions are discussed for the strong-coupling
regime of the Higgs sector.

\section{Abelian $U(1)$ theory}
In the standard model of electroweak interactions the physical vacuum is
believed to be at a broken symmetry minimum of the effective potential.
Since the $SU(2)\times U(1)$ gauge symmetry 
is broken to the electromagnetic $U(1)$ group, the full gauge invariance of the
GEP is not a real issue, provided that the method is shown to be
a genuine variational calculation.
However the electromagnetic $U(1)$ symmetry must remain unbroken in the
physical vacuum and the method must give the same predictions for
any choice of the unbroken $U(1)$ gauge.
In this section we discuss the Abelian $U(1)$ theory
and show how the GEP can be made invariant by
a simple optimization of the variational approximation.

The GEP for the Abelian $U(1)$ theory (scalar electrodynamics) has been discussed
by several authors\cite{ibanez,camarda}, and has been recently shown to provide
a general interpolation of the experimental data for superconductors\cite{marotta}.
We briefly review the method in order to discuss its gauge invariance.
In the Euclidean formalism the action reads
\BE
S=\int dx \left[\frac{1}{4} F_{\mu \nu}F^{\mu
\nu}+\frac{1}{2}(D_{\mu}\phi)^{*}(D^{\mu}\phi)+
\frac{1}{2}m^{2} \phi^{*}\phi+
\lambda(\phi^{*}\phi)^{2}\right].
\label{action1}
\EE
where $\phi$ is a complex (charged) scalar field, its covariant
derivative is defined according to
\BE
D_{\mu} = \partial_{\mu}+ie A_{\mu}
\label{derivative}
\EE
and 
$F_{\mu\nu}=\partial_\mu A_\nu-\partial_\nu A_\mu$.
The vacuum to vacuum transition amplitude is written as the functional integral
\BE
Z=\int D[\phi,\phi^*,A_\mu]e^{-S}.
\label{z}
\EE
In four space-time dimensions the action can be regarded as a toy model for 
the Higgs sector of standard electroweak interactions.
In three dimensions the same action gives the static Ginzburg-Landau
free-energy of a superconductor and $Z$ plays the role of the partition function.

The action $S$ in Eq.(\ref{action1}) has a local $U(1)$ symmetry as it is invariant
for a local gauge transformation 
\BE
{A_\mu} \to
{A_\mu}+{\partial_\mu} \chi(x)
\EE
\BE
\phi\to\phi\> e^{\displaystyle{-ie\chi(x)}}
\EE
where $\chi(x)$ is an arbitrary function. 
The integration over $A_\mu$ is then redundant in Eq.(\ref{z}) and
a gauge fixing term must be inserted 
according to the standard De Witt-Faddeev-Popov method\cite{dewitt} 
\BE
Z=\int D[\phi,\phi^*,A_\mu]e^{-S}e^{-S_{fix}}.
\label{zfix}
\EE
where the gauge fixing action is
\BE
S_{fix}=\int d x \frac{1}{2\epsilon}f^2
\label{sfix}
\EE
and $f(A,\phi)=0$ is an arbitrary gauge constraint. 

$Z$ is invariant for any change of the parameter $\epsilon$ and
of  the constraint $f$. With some abuse of language, this invariance property 
is referred to as gauge
invariance while it is a more general invariance since $Z$ does not 
depend on the shape of the weight factor which
has been added in Eq.(\ref{zfix}). 
Only for $\epsilon\to 0$ the weight factor $\exp(-S_{fix})$ becomes a 
$\delta$-function which enforces
the constraint $f=0$ on the fields.  
Thus standard gauge invariance denotes the invariance of
the theory for any change of the constraint $f=0$ in the limit $\epsilon\to 0$. 
This is a weaker condition, 
but unfortunately even this is not fulfilled by some approximations.
In this paper we will make a distinction between the generalized gauge invariance
and the standard gauge invariance, since it turns out that Jensen's inequality
may break the first while leaving the second unbroken.

In fact let us take $f=(\partial_\mu A^\mu)$ and switch to polar coordinates  
$\phi\to\rho\exp(i\gamma)$ in the
functional integral. The amplitude $Z$ reads
\BE
Z=\int D[A_\mu, \rho^2] e^{\displaystyle{-\int dx {\cal L}}}
\int D[\gamma] e^{\displaystyle{-\int d x{\cal L}_\gamma}}
\label{zpol}
\EE
where ${\cal L}$ is the phase independent Lagrangian 
\BE
{\cal L}=\frac{1}{4} F_{\mu \nu}F^{\mu
\nu}
+\frac{1}{2}\partial_{\mu}\rho\partial^{\mu}\rho+
\frac{1}{2}m^{2}\rho^2+
\lambda\rho^{4}+\frac{1}{2}e^2\rho^2{A_\mu A^\mu}+
\frac{1}{2\epsilon} (\partial_\mu A^\mu)^2
\label{L}
\EE
and ${\cal L}_{\gamma}$ is the sum of the Lagrangian terms which 
depend on the phase $\gamma$
\BE
{\cal L}_\gamma=
\frac{1}{2}\rho^2\partial_{\mu}\gamma\partial^{\mu}\gamma+
e\rho^2\partial _\mu\gamma A^\mu.
\label{Lgamma}
\EE

As a first step towards a gauge invariant GEP the phase $\gamma$ 
is integrated by use of Jensen's inequality. While this integration
has been sometimes regarded as exact\cite{camarda,kleinert} it is not, but
can be shown to be a genuine variational approximation\cite{marotta}.
In order to show that, let us denote by ${\cal L}_0$ 
the first term in the phase dependent Lagrangian Eq.(\ref{Lgamma})
\BE
{\cal L}_0={\cal L}_\gamma(e=0)=
\frac{1}{2}\rho^2\partial_{\mu}\gamma\partial^{\mu}\gamma.
\label{L0}
\EE
We observe that up to constant factors the exact integration over $\gamma$ yields
\BE
\int D[\gamma] e^{\displaystyle{-\int d x{\cal L}_0}}\sim\prod_x \frac{1}{\rho}.
\label{factor}
\EE
Thus we may write the $D[\rho^2]$ integral in Eq.(\ref{zpol}) as
\BE
Z\sim\int D[A_\mu, \rho] e^{{-\int dx {\cal L}}}\>
\left\{ 
\frac 
{\int D[\gamma] e^{{-\int d x{\cal L}_\gamma}} }
{\int D[\gamma] e^{{-\int d x{\cal L}_0}} }  
\right\}.
\label{zav}
\EE
An average over the phase can be defined as
\BE
\langle(...)\rangle_\gamma=
\frac
{\int D[\gamma] e^{-\int d x{\cal L}_0}(...)} 
{\int D[\gamma] e^{-\int d x{\cal L}_0}}  
\label{average}.
\EE
and with this notation the exact $Z$ amplitude Eq.(\ref{zpol}) reads
\BE
Z=\int D[A_\mu, \rho] e^{\displaystyle{-\int dx {\cal L}}}
\left\langle e^{\displaystyle{-\int d x e\rho^2\partial_\mu \gamma A^\mu}}
\right\rangle_\gamma.
\label{zav2}
\EE
Then the convexity of the exponential function (Jensen's inequality)
ensures that
\BE
Z\geq \int D[A_\mu, \rho] \>e^{\displaystyle{-\int dx {\cal L}}}
\> e^{\displaystyle{-\int \left\langle e\rho^2 \partial _\mu\gamma 
A ^\mu\right\rangle_\gamma d x}}.
\label{diseq}
\EE
The average in the right hand side vanishes (it is linear in $\gamma$), and
the approximate amplitude 
\BE
Z_p=\int D[A_\mu, \rho] e^{\displaystyle{-\int dx {\cal L}}}
\label{zappr}
\EE
satisfies the variational constraint
\BE
Z\geq Z_p
\EE
so that the approximate potential ${\cal V}_p=-\ln Z_p$ 
is bounded by the exact one
${\cal V}=-\ln Z$ 
\BE
{\cal V}_p\geq {\cal V}.
\label{bound}
\EE
Eventually
the GEP may be evaluated
by the same $\delta$ expansion method discussed in Ref.\cite{ibanez} and
\cite{stancu} and also reported by Ref.\cite{camarda}. 
Inserting a source term for the real field $\rho$ the generating functional $Z_p[J]$ reads
\BE
Z_p[J]=\int D[A_\mu, \rho] \>e^{\displaystyle{-\int dx {\cal L}}}
\>e^{\displaystyle{-\int dx J\rho}}
\label{zj}
\EE
and the effective potential is recovered by the Legendre transform
\BE
{\cal V}_p[\varphi]=-\ln Z_p+\int d x J\varphi
\label{v}
\EE
where $\varphi$ is the average value of $\rho$.
While the details of the derivation may be found in Ref.\cite{marotta},
we would like to discuss the gauge invariance of the result and the effects
of any gauge change on the approximation. 

Although Eq.(\ref{bound}) ensures that
the approximate effective potential is bounded by the exact effective potential,
a different choice of gauge could change the approximate result. Let us take in
account the effects of a gauge change on the approximation in Eq.(\ref{diseq}).
A generic gauge change can be introduced by a shift of the constraint 
$f\to f^\prime\not= \partial_\mu A^\mu$. After that we can always restore the
constraint by a simple change of variables 
$A^\mu \to A^\mu+\partial_\mu \Lambda$ in the functional integration:
the function $\Lambda$ can be chosen in order to make $f^\prime=\partial_\mu A^\mu$
again. However the change of variables adds some extra terms to the Lagrangian.
Some of them only add constant or vanishing contributions (up to surface terms).
Some others do not cancel but they only change the phase dependent Lagrangian
${\cal L}_\gamma$ which becomes
\BE
{\cal L}_\gamma=
\frac{1}{2}\rho^2
\left(\partial_{\mu}\gamma+e\partial^{\mu}\Lambda\right)^2+
e\rho^2\left(\partial _\mu\gamma +e\partial_\mu \Lambda\right)A^\mu.
\label{Lgamma2}
\EE
It is obvious that we are allowed to restore the former approximate result by a simple
shift of the phase $\gamma\to \gamma-e\Lambda$ before integrating. There
is nothing wrong in that change as it is just an other change of variables in the
functional  integral. However we might want to keep the extra terms in
${\cal L}_\gamma$ and approximate the generating functional through Jensen's
inequality according to Eq.(\ref{diseq}). In that case we would get a different
result as extra terms would remain in the Lagrangian. In other words the
variational approximation seems to depend on our choice for the integration
variable $\gamma$.

In fact the same ambiguity arises whenever we approximate a simple Gaussian integral
by Jensen's inequality. For instance let us consider the exact result
\BE
I=\int_{-\infty}^\infty e^{-bx^2+ax} dx=\sqrt{\frac {\pi}{b}} e^{\frac{a^2}{4b}}
\EE
where $a$ and $b>0$ are arbitrary real parameters. According to
Eq.(\ref{average}) we can define
the average 
\BE
\langle(...)\rangle_b=
\frac
{\int_{-\infty}^\infty e^{-bx^2}(...)} 
{\int_{-\infty}^\infty e^{-bx^2}}  
\label{averageb}.
\EE
The integral can be approximated by use of Jensen's inequality
\BE
I=\int_{-\infty}^\infty e^{-bx^2} \langle e^{ax}\rangle_b dx
\ge\sqrt{\frac {\pi}{b}}
e^{\langle{ax}\rangle_b}=\sqrt{\frac {\pi}{b}}.
\EE
In fact this approximate result is smaller than the exact one, 
but it can be improved by the change of variable $x\to x+y$:
\BE
I=e^{-by^2+ay}\int_{-\infty}^\infty e^{-bx^2} \langle e^{(a-2by)x}\rangle_b dx
\EE
Of course the exact result does not change, but the approximate estimate does
depend on y:
\BE
I\ge e^{-by^2+ay}\sqrt{\frac {\pi}{b}}e^{\langle{(a-2by)x}\rangle_b}=
\sqrt{\frac {\pi}{b}}e^{-by^2+ay}.
\EE
The exponent $(-by^2+ay)$ has a maximum at the saddle point 
$y=a/(2b)$ where the approximate
estimate reaches the exact result. 
It emerges that, in order to improve our variational method, the integration
variable $\gamma$ must be shifted before the use of Jensen's inequality.
The best phase shift $\delta\gamma$ should be the saddle point
of the modified ${\cal L}_\gamma$ Eq.(\ref{Lgamma2}), and should
satisfy the linear equation
\BE
\partial_\mu (\delta\gamma +e\Lambda)=-eA_\mu.
\EE
Unfortunately in general there is no solution unless $A^\mu$ is a pure gauge.
And this is the reason why phase integration does not give an exact result as
it was claimed\cite{camarda,kleinert}. The best we can do is to require
that the phase change cancels the longitudinal component of the gauge field
$e\partial_\mu \Lambda$ which has been inserted by the gauge change. That
is equivalent to take $\delta\gamma=-e\Lambda$ which is exactly the same
choice required in order to cancel all the effects of the gauge change in 
the Lagrangian ${\cal L}_\gamma$ Eq.(\ref{Lgamma2}). 

Thus an optimized use of Jensen's inequality yields the very same approximate
result for any choice of the gauge. In this sense we may state that the GEP
obtained by this method is fully gauge invariant.

The whole discussion only makes sense in the limit $\epsilon\to 0$ when
the gauge fixing term enforces the constraint $f=0$ on the fields. For a general
choice of $\epsilon$ there is no fixed gauge and the functional integration
runs over the gauge group. This further averaging over the gauge group worsen
the approximate result obtained by Jensen's inequality, and the result turns
out to depend on $\epsilon$. Of course the best choice is the limit $\epsilon\to 0$
since in that case the integration over the gauge group does not introduce 
any further approximation through the inequality which becomes exact.

\section{Non Abelian $SU(2)\times U(1)$ theory}

In this section we discuss the variational method for the full
$SU(2)\times U(1)$ gauge group of electroweak interactions.
The theory is described by the Euclidean Lagrangian
\BE
{\cal L}=
\frac{1}{2}(D_{\mu}\Phi)^{\dagger}(D^{\mu}\Phi)+V(\Phi^\dagger\Phi)+
{\cal L}_{YM}
\label{Lsu2}
\EE
where $\vec A_\mu$, $B_\mu$ are the gauge fields, ${\cal L}_{YM}$ is
the Yang-Mill Lagrangian
\BE
{\cal L}_{YM}=\frac{1}{4} \vec F_{\mu \nu}\cdot \vec F^{\mu\nu}+
\frac{1}{4}\left( \partial_\mu B_\nu-\partial_\nu B_\mu\right)^2
\label{LYM}
\EE
in terms of the fields
\BE
\vec F_{\mu\nu}=\left( \partial_\mu \vec A_\nu-\partial_\nu \vec A_\mu\right)
+g\vec A_\mu\times\vec A_\nu
\EE
and $\Phi$ is a Higgs doublet of complex fields $\phi_1$, $\phi_2$
\BE
\Phi=
\left(\begin{array}{c}
\phi_1\\
\phi_2\\
\end{array}\right).
\EE
The covariant derivative reads
\BE
D_\mu=\left[\partial_\mu -ig \vec A_\mu\cdot \vec {\bf T}
+i g^\prime B_\mu \frac{\bf Y}{2}\right]
\EE
where $g$, $g^\prime$ are the weak couplings and the generators are defined
by the $2\times 2$ matrices ${\bf Y}=1$ and $\vec {\bf T}=\frac{1}{2}\vec \sigma$.
As usual the charge operator is ${\bf Q}=e({\bf T}_3+{\bf Y}/2)$.

In general the Higgs doublet $\Phi$ can be parametrized according to
\BE
\Phi=\rho e^{i\gamma} e^{i \sigma_3 \phi}
\left(\begin{array}{c}
\cos\theta\\
\sin\theta\\
\end{array}\right)
\EE
where $\rho\ge 0$ is a real field, and the three phases $\gamma$, $\phi$, $\theta$
may be taken in the ranges $0\le\gamma\le 2\pi$, 
$0\le\phi\le 2\pi$ and  $0\le\theta\le \pi/2$. 
Without fixing any special gauge we would like to discuss some general properties
of the generating functional
\BE
Z[J]=\int D[\phi_1,\phi_2,\vec A,B] \>e^{\displaystyle{-\int d^4x 
\left({\cal L}-\rho J\right)}}.
\label{zsu2}
\EE
A change of integration variables yields
\BE
Z[J]=\int D[\rho^4,\sin^2\theta,\gamma,\phi,\vec A,B] 
\>e^{\displaystyle{-\int d^4x 
\left({\cal L}-\rho J\right)}}
\label{zsu2ph}
\EE
where ${\cal L}$ can be written as
\BE
{\cal L}={\cal L}_\rho+{\cal L_G}+{\cal L}_1+{\cal L}_2+{\cal L}_{YM}
\label{Ltot}
\EE
according to the following definitions:
${\cal L}_\rho$ is the Lagrangian of the self-interacting scalar real field $\rho$
\BE
{\cal L}_\rho=\frac{1}{2}(\partial_\mu\rho)^2+V(\rho^2);
\label{Lrho}
\EE
${\cal L_G}$ contains the gauge phase quadratic terms
\BE
{\cal L_G}=\frac{1}{2}\rho^2 \left[
(\partial_\mu\gamma)^2+(\partial_\mu\phi)^2+
(\partial_\mu\theta)^2\right];
\label{LG}
\EE
${\cal L}_2$ contains quadratic interaction terms for the gauge fields
\BE
{\cal L}_2=\frac{1}{8}\rho^2 \left[
g^2\vec A_\mu\cdot \vec A^\mu
-2 g g^\prime B^\mu \vec A_\mu\cdot \vec R
+{g^\prime}^2B_\mu B^\mu\right];
\label{L2}
\EE
where $\vec R$ is the phase dependent vector
\BE
\vec R=\left(\begin{array}{c}
\sin (2\theta) \cos (2\phi)\\
-\sin(2\theta)\sin (2\phi)\\
\cos(2\theta)\\
\end{array}\right);
\EE
${\cal L}_1$ contains linear interaction terms for the gauge fields 
\BE
{\cal L}_1=
\frac{1}{2}\rho^2 g \vec A^\mu\cdot \vec\Gamma_\mu
+\frac{1}{2}\rho^2 g^\prime  B^\mu \Theta_\mu
\label{L1}
\EE
where $\vec\Gamma_\mu$ and $\Theta_\mu$ depend on
phases and are defined as follows

\BE
\vec\Gamma_\mu=\vec R\partial_\mu \gamma+\left(\begin{array}{c}
\sin (2\phi)\partial_\mu\theta\\
\cos (2\phi)\partial_\mu\theta\\
-\partial_\mu\phi\\
\end{array}\right)
\EE
\BE
\Theta_\mu=\partial_\mu\gamma+\cos(2\theta)\partial_\mu\phi.
\EE
According to the standard De Witt-Faddeev-Popov 
method\cite{dewitt} the integration over the gauge group
can be dealt with by insertion of a gauge fixing term 
\BE
{\cal L}_{fix}=-\frac{1}{\epsilon} (f_\alpha)^2
\EE
where the index $\alpha$ runs over the four gauge fields
\BE
f_\alpha=(\vec f, f_{B}).
\EE
The gauge invariance of the generating
functional $Z[J]$ is preserved provided that a factor is also inserted
in the integrand, equal to the determinant of the matrix
\BE
{\cal F}_{\alpha,\beta}=
\left(\frac {\delta f_\alpha}{\delta \lambda_\beta}\right)_{\lambda_\beta=0}
\EE
where $\lambda_\beta$ is the generic parameter 
of a gauge transformation\cite{weinberg2}.
The gauge invariant generating functional now reads
\BE
Z[J]=
\int D[\rho^4,\sin^2\theta,\gamma,\phi,\vec A,B]\> 
{\rm Det}{\cal F}
\>e^{\displaystyle{-\int d^4x 
\left({\cal L+L}_{fix}-\rho J\right)}}.
\label{zsu2fix}
\EE

From a formal point of view the determinant can be seen as
\BE
{\rm Det}{\cal F}=e^{\displaystyle{-\int d^4x {\cal L}_{gh}}}
\EE
where the {\it ghost} Lagrangian ${\cal L}_{gh}$ 
\BE
{\cal L}_{gh}=-{\rm Tr}\log {\cal F}
\label{Lgh}
\EE
can be written in terms of anticommuting Grassmann ghost fields.
Thus the definition of the generating functional $Z[J]$ in
Eq.(\ref{zsu2}) can be made gauge invariant by the replacement
${\cal L}\to{\cal L}+{\cal L}_{fix}+{\cal L}_{gh}$.

We would like to extend the variational method discussed in the previous
section, and see how it works for the non-Abelian model.
In analogy to Eq.(\ref{factor}) we can see that integration over phases
yields
\BE
\int D[\sin^2\theta,\gamma,\phi] 
e^{\displaystyle{-\int d^4x{\cal L_G}}}\sim\prod_x \frac{1}{\rho^3}.
\label{factor3}
\EE
Let us use the shorthand notation $D_\gamma=D[\sin^2\theta,\gamma,\phi]$
and $D_\rho=D[\rho,\vec A,B]$, and define the average over phases
\BE
\langle(...)\rangle_\gamma=
\frac
{\int D_\gamma e^{-\int d^4 x{\cal L_G}}(...)} 
{\int D_\gamma e^{-\int d^4 x{\cal L_G}}}  
\label{average3}.
\EE
The generating functional then reads
\BE
Z[J]=\int D_\rho \>e^{\>\displaystyle{\int d^4x \rho J}}\>
\left\langle e^{\displaystyle{-\int d^4x ({\cal L}+{\cal L}_{fix}
+{\cal L}_{gh}-{\cal L_G})}}
\right\rangle_\gamma.
\label{zav3}
\EE
Moreover for any trial Gaussian Lagrangian ${\cal L}_{GEP}(\rho,\vec A,B)$
which does not depend on the phases $\theta$, $\phi$, $\gamma$, a
further average can be defined 
\BE
\langle(...)\rangle_\rho=
\frac
{\int D_\rho e^{-\int d^4 x{\cal L}_{GEP}}(...)} 
{\int D_\rho e^{-\int d^4 x{\cal L}_{GEP}}}  
\label{avGEP}.
\EE
and the exact gauge invariant generating functional can be written as a
double average
\BE
Z[J]=
\left\langle
e^{\>\displaystyle{\int d^4x \rho J}}\>
\left\langle e^{\displaystyle{-\int d^4x ({\cal L}+{\cal L}_{fix}
+{\cal L}_{gh}-{\cal L_G}-{\cal L}_{GEP})}}
\right\rangle_\gamma\>\right\rangle_\rho Z_0
\label{zavGEP}
\EE
where 
\BE
Z_0=\int D_\rho e^{\>\displaystyle{-\int d^4x {\cal L}_{GEP}}}.
\EE

A variational approximation for the effective potential follows from
the use of Jensen's inequality: the approximate generating functional
$Z_{GEP}[J]$ is bound by the exact one as
\BE
Z[J]\ge Z_{GEP}[J]=Z_0 \>
e^{\displaystyle{-\int d^4x 
\left\langle\left\langle{\cal L}+{\cal L}_{fix}
+{\cal L}_{gh}-{\cal L_G}-{\cal L}_{GEP}-J\rho
\right\rangle_\gamma\right\rangle_\rho}}.
\label{Jens1}
\EE
Up to a total volume factor, 
the exact effective potential is defined as the Legendre transform
\BE
{\cal V}[\bar\rho]=-\log Z[J]+\int d^4x J\bar \rho
\EE
where $\bar \rho$ is the expectation value of the field $\rho$ in
the presence of the source $J$.
We assume that $\langle\rho\rangle_\rho=\bar\rho$ where $\bar\rho$ is
a parameter of the trial Lagrangian ${\cal L}_{GEP}$. In other words
$\bar\rho$ is the central value of the quadratic Lagrangian ${\cal L}_{GEP}$.
It follows that 
\BE
{\cal V}[\bar\rho]\le{\cal V}_{GEP} (\bar\rho)=
-\log Z_{GEP}[J]+\int d^4x J\bar \rho
\label{Vtrue}
\EE
Thus the approximate Gaussian effective potential ${\cal V}_{GEP}$ is a
genuine variational approximation of
the exact effective potential, and can be evaluated by the double
average 
\BE
{\cal V}_{GEP} (\bar\rho)=
-\log\int D_\rho e^{\>\displaystyle{-\int d^4x {\cal L}_{GEP}}}
+\int d^4x \left\langle\left\langle{\cal L}+{\cal L}_{fix}
+{\cal L}_{gh}-{\cal L_G}-{\cal L}_{GEP}
\right\rangle_\gamma\right\rangle_\rho.
\label{VGEP}
\EE
The present derivation holds for any gauge choice, that means the method can be
improved by a gauge change. In fact, as for the Abelian $U(1)$ theory, the limit
$\epsilon\to 0$ should be imposed on ${\cal L}_{fix}$ in order to improve the
reliability of Jensen's inequality in Eq.(\ref{Jens1}). Under that limit the
integration over the gauge group does not introduce new approximations as the
constraint in ${\cal L}_{fix}$ yields a $\delta$ function and the integration
over the gauge group becomes exact (it is not affected by the inequality).
On the other hand a gauge choice should not be a problem as the gauge symmetry
is broken anyway in the physical vacuum.

The physics of the non-Abelian $SU(2)\times U(1)$ model is more evident in unitarity
gauge which seems to be the natural choice for dicussing the symmetry breaking
mechanism. However there is a more formal motivation for that choice which
has to do with calculability. Provided that we take a simple quadratic shape
for the trial Lagrangian ${\cal L}_{GEP}$, the Gaussian integral and
the averages in Eq.(\ref{VGEP}) can be all easily evaluated with the important
exception of the ghost term $\langle {\cal L}_{gh}\rangle$. The existence of this
term makes the method useless since we do not know how to calculate its average.
In a naive approach we could write ${\cal L}_{gh}$ in terms of anticommuting
Grassmann ghost fields, but Jensen's inequality cannot be proven for Grassmann
variables and the result would not be a genuine variational approximation.
There would be no control on the approximation. An other naive approach would
consist in the mere neglection of this term, and that can be shown to be
the tree-level approximation of a perturbative expansion.

However in unitarity gauge the constraint functions $f_\alpha$ do not depend
on the gauge fields: the mass of the ghost fields scales like 
$\epsilon^{-1/2}$ and becomes infinite in the $\epsilon\to 0$ limit,
decoupling the ghosts from the physical fields. 
In other words the factor ${\rm Det}{\cal F}$ in Eq.(\ref{zsu2fix}) 
becomes a constant and can be carried out of the integral. Thus in unitarity
gauge the average of ${\cal L}_{gh}$ is a constant and can be neglected.
We conclude that
calculability makes the choice of unitarity gauge the only viable choice.

It is instructive to study the behaviour of ${\cal L}_{gh}$ in the
renormalizable $\xi$-gauge of Fujikawa, Lee and Sanda\cite{fujikawa}
which is equivalent to the unitarity gauge in the $\epsilon=1/\xi\to 0$ limit.
The matrix ${\cal F}$ can be written as\cite{weinberg2}
\BE
{\cal F}={\cal F}_0+{\cal F}_{int}
\EE
where ${\cal F}_{int}$ contains a linear coupling with the gauge fields,
${\cal F}_0$ is the matrix
\BE
({\cal F}_0)_{\alpha x,\beta y}=\left[\delta_{\alpha\beta}
\partial_\mu\partial^\mu+\frac{1}{\epsilon} M_{\alpha\beta}\right]
\delta^4 (x-y)
\label{invprop}
\EE
and $M_{\alpha\beta}$ is a constant mass matrix. Inserting
the definition Eq.(\ref{Lgh}) in Eq.(\ref{VGEP}),
the double average of ${\cal L}_{gh}$  can be written as
\BE
\left\langle\left\langle{\cal L}_{gh}\right\rangle\right\rangle=
-\left\langle\left\langle {\rm Tr}\log {\cal F}_0\right\rangle\right\rangle
-\left\langle\left\langle {\rm Tr}
\log(1+{\cal F}_0^{-1} {\cal F}_{int})\right\rangle\right\rangle
\EE
The second term can be expanded yielding the perturbative series
\BE
\left\langle\left\langle {\rm Tr}
\log(1+{\cal F}_0^{-1} {\cal F}_{int})\right\rangle\right\rangle
\approx{\rm Tr}
\left\langle\left\langle 
{\cal F}_0^{-1} {\cal F}_{int}\right\rangle\right\rangle
-\frac{1}{2}{\rm Tr}
\left\langle\left\langle 
{\cal F}_0^{-1} {\cal F}_{int}{\cal F}_0^{-1} {\cal F}_{int}
\right\rangle\right\rangle+\dots
\label{expans}
\EE
According to Eq.(\ref{invprop})
the average $\langle\langle{\cal F}_0^{-1}\rangle\rangle$ can
be regarded as the propagator for a massive particle 
(a ghost) whose
mass scales like $1/\sqrt{\epsilon}$. The interaction vertex
${\cal F}_{int}$ is linear in the gauge fields, and the average
of any pair 
$\langle\langle{\cal F}_{int}{\cal F}_{int}\rangle\rangle$
yields a gauge field propagator.
Thus a diagrammatic expansion is recovered by Wick's theorem:
Eq.(\ref{expans}) can be regarded as the sum of loop
diagrams each consisting of a closed ghost
ring crossed by any number of gauge lines.
At tree-level, neglecting all the interaction lines, 
the double average of ${\cal L}_{gh}$ becomes a constant and
can be neglected in the effective potential Eq.(\ref{VGEP}).
Thus the naive neglection of ${\cal L}_{gh}$ is equivalent
to the tree-level approximation of the perturbative expansion.
However, in the $\epsilon\to 0$ limit, the ghost mass becomes
infinite and all the terms in the expansion vanish.
In the $\epsilon\to 0$ limit the renormalizable $\xi$-gauge
becomes the unitarity gauge, and we recover the result that
${\cal L}_{gh}$ can only be neglected in the unitarity gauge.

With that gauge choice understood, the double average in 
Eq.(\ref{VGEP}) becomes trivial and the GEP can
be easily evaluated provided that a simple quadratic shape
is chosen for ${\cal L}_{GEP}$. Moreover if ${\cal L}_{GEP}$ is an
even functional the double average of ${\cal L}_1$ also vanishes.
However, we have seen in the previous section that, in order
to get the best approximation from Jensen's inequality,
the linear term must be shifted. In the Abelian $U(1)$ model
the best choice was the transverse gauge fixed by the constraint
$\partial_\mu A^\mu=0$. In Unitarity gauge we still have a free
overall electromagnetic $U(1)$ phase, and the best approximation
arises from the transverse electromagnetic gauge.
In order to show that, we must take a shift of the integration variables
before taking the average. A linear change of variables is required first
from the gauge fields $\vec A_\mu$, $B_\mu$ to the physical fields
$W_\mu^{\pm}$, $Z_\mu$, $A_\mu$; then the best shift for the electromagnetic
phase can be discussed, and eventually the double average will be taken.

\section{GEP for the Standard Model}

In the unitarity gauge ($\theta=\pi/2$) the physical massive 
gauge fields $W^{\pm}$, $Z$ and the electromagnetic gauge field $A$ are
defined according to the linear transformation
\BE
A^1_\mu=\frac{W^+_\mu+W^-_\mu}{\sqrt{2}}
\label{tran1}
\EE
\BE
A^2_\mu=\frac{W^+_\mu-W^-_\mu}{i\sqrt{2}}
\label{tran2}
\EE
\BE
A^3_\mu=\frac{e}{g}A_\mu-\frac{e}{g^\prime} Z_\mu
\label{tran3}
\EE
\BE
B_\mu=-\frac{e}{g}Z_\mu-\frac{e}{g^\prime} A_\mu
\label{tran4}
\EE
where the electromagnetic charge $e$ follows from the
constraint
\BE
\frac{e^2}{g^2}+\frac{e^2}{{g^\prime}^2}=1.
\label{charge}
\EE
Insertion of these definitions in the quadratic Lagrangian term Eq.(\ref{L2}) 
yields
\BE
{\cal L}_2=\frac{\rho^2}{v^2} M_W^2 W^+_\mu{W^-}^\mu
+\frac{1}{2}\frac{\rho^2}{v^2} M_Z^2 Z_\mu Z^\mu
\label{LSM2}
\EE
where the masses $M_W$ and $M_Z$ are given by the standard definitions
\BE
M_W=\frac{vg}{2}
\label{MW}
\EE
\BE
M_Z=\frac{1}{2} v \sqrt{g^2+{g^\prime}^2}
\label{MZ}
\EE
in terms of the free parameter $v$.
The gauge field $A_\mu$ remains massless, as it must be, since the electromagnetic
$U(1)$ symmetry is unbroken. According to
the discussion of Section II, for the $U(1)$ gauge group we get 
the best variational approximation 
in the transverse gauge $\partial_\mu A^\mu=0$. That constraint is imposed
by still taking the gauge-fixing term to be
\BE
{\cal L}_{fix}=\frac{1}{\epsilon} \left(\partial_\mu A^\mu\right)^2
\EE
where the limit $\epsilon\to 0$ is understood.
This gauge choice is equivalent to a shift of the integration variables before
the average, in order to cancel the longitudinal part of the gauge field $A_\mu$.
Then the average can be taken in Eq.(\ref{VGEP}) and, provided that ${\cal L}_{GEP}$
is even, the odd lagrangian terms give a vanishing contribution. Thus we can drop
${\cal L}_1$ and the odd terms of ${\cal L}_{YM}$ in the average, and the ghost
term ${\cal L}_{gh}$ which does not contribute in the unitarity gauge.
Insertion of Eq.(\ref{Ltot}) in the effective potential Eq.(\ref{VGEP}) yields 
\BE
{\cal V}_{GEP} (\bar\rho)=
-\log\int D_\rho e^{\>\displaystyle{-\int d^4x {\cal L}_{GEP}}}
+\int d^4x 
\left\langle{\cal L}_{int}
\right\rangle_\rho.
\label{VGEP2}
\EE
where the interaction Lagrangian now reads
\BE
{\cal L}_{int}={\cal L}_\rho+{\cal L}_{fix}
+{\cal L}_2+{\cal L}_{YM}^{even}-{\cal L}_{GEP}
\label{Lint}
\EE
Next we take a shift of the scalar field $\rho$, and as usual\cite{stancu}
we define the scalar Higgs field $h$ according to 
\BE
h=\rho-\bar\rho.
\label{higgs}
\EE
A natural choice for the Gaussian trial Lagrangian is the sum
of quadratic Gaussian Lagrangians for the gauge fields and the
scalar Higgs field
\BE
{\cal L}_{GEP}={\cal L}_{GEP}(h)+{\cal L}_{GEP}(W)+{\cal L}_{GEP}(Z)+{\cal L}_{GEP}(A)
\label{LGEPTOT}
\EE
with the Lagrangian terms defined according to
\BE
{\cal L}_{GEP}(h)=
\frac{1}{2} \left(\partial_\mu h\right)^2+
\frac{1}{2} \Omega_h^2 h^2
\label{LGEPh}
\EE
\BE
{\cal L}_{GEP}(W)=
\frac{1}{2} \left(\partial_\mu W^+_\nu-\partial_\nu W^+_\mu\right)
\left(\partial_\mu W^-_\nu-\partial_\nu W^-_\mu\right)
+\Omega_W^2 W^+_\mu {W^-}^\mu
\label{LGEPW}
\EE
\BE
{\cal L}_{GEP}(Z)=
\frac{1}{4} \left(\partial_\mu Z_\nu-\partial_\nu Z_\mu\right)^2
+\frac{1}{2} \Omega_Z^2 Z_\mu Z^\mu
\label{LGEPZ}
\EE
\BE
{\cal L}_{GEP}(A)=
\frac{1}{4} \left(\partial_\mu A_\nu-\partial_\nu A_\mu\right)^2
+\frac{1}{\epsilon} \left(\partial_\mu A^\mu\right)^2.
\label{LGEPA}
\EE
Here the masses $\Omega_h$, $\Omega_W$ and $\Omega_Z$ must be regarded
as variational parameters. 
With this choice we get $\langle h\rangle=0$
and by the definition of $h$, Eq.(\ref{higgs}), then $\langle \rho\rangle=\bar\rho$
as it was required in the derivation of the Gaussian 
effective potential Eq.(\ref{VGEP}).
In order to evaluate ${\cal V}_{GEP} (\bar\rho)$, according to Eq.(\ref{VGEP2})
we also need ${\cal L}_{int}$ which now reads
\BA
{\cal L}_{int}&=&V\left( (\bar \rho+h)^2\right)-\frac{1}{2}\Omega_h^2 h^2+
\nonumber\\
&+&\left[\left(\frac{\bar\rho+h}{v}\right)^2 M_W^2-\Omega_W^2\right]
W^+_\mu {W^-}^\mu+
\frac{1}{2}
\left[\left(\frac{\bar\rho+h}{v}\right)^2 M_Z^2-\Omega_Z^2\right]
Z_\mu Z^\mu+{\cal L}_4
\label{Lint2}
\EA
where ${\cal L}_4$ contains the quartic terms that come out from 
the product $(\vec A_\mu\times\vec A_\nu)^2$ in ${\cal L}_{YM}^{even}$
\BA
{\cal L}_4&=&e^2\left[(A_\mu A^\mu)(W^+_\nu {W^-}^\nu)
-(W^+_\mu A^\mu)(W^-_\nu A^\nu)\right]
+e^2\frac{g^2}{{g^\prime}^2}
\left[(Z_\mu Z^\mu)(W^+_\nu {W^-}^\nu)
-(W^+_\mu Z^\mu)(W^-_\nu Z^\nu)\right]+
\nonumber\\
&+&e^2\frac{g}{{g^\prime}}
\left[(W^+_\mu A^\mu)(W^-_\nu Z^\nu)
-2(A_\mu Z^\mu)(W^+_\nu {W^-}^\nu)+
(W^+_\mu Z^\mu)(W^-_\nu A^\nu)\right]+
\nonumber\\
&+&\frac{1}{2}g^2 \left[
(W^+_\mu {W^-}^\mu)^2
-(W^+_\mu {W^+}^\mu)(W^-_\nu {W^-}^\nu)\right].
\label{L4}
\EA
The couplings can be written in terms of the mass parameters by the
standard relations
\BE
g^2=\frac{4 M_W^2}{v^2}
\EE
\BE
e^2=\frac{4 M_W^2}{v^2}\left(1-\frac{M_W^2}{M_Z^2}\right)
\EE
\BE
\frac{g}{g^\prime}=\frac{M_W}{\sqrt{M_Z^2-M_W^2}}.
\EE
However at this stage $M_W$ and $M_Z$ are just an alternative
set of parameters and they are not physical masses.
 
The explicit evaluation of the Gaussian effective potential then follows
by Wick's theorem through Eq.(\ref{VGEP2}). 
As usual, the classical potential of the standard Higgs sector is
written as
\BE
V(\rho^2)=\frac{1}{2}m^2 \rho^2+\frac{1}{4!}\lambda \rho^4
\label{Vclassical}
\EE
and denoting by $\varphi$ the average of the field $\rho$, $\varphi=\bar \rho$, 
a straightforward calculation yields
the effective potential (GEP) 
\BA
{\cal V}_{GEP} (\varphi)&=&\frac{1}{2} m^2\varphi^2+\frac{1}{2}m^2 I_0(\Omega_h)
+\frac{\lambda}{4!}\varphi^4+\frac{\lambda}{4}\varphi^2 I_0(\Omega_h)+
\frac{\lambda}{8}\left[ I_0(\Omega_h)\right]^2-\frac{1}{2}\Omega_h^2 I_0(\Omega_h)+
\nonumber\\
&+&I_1(\Omega_h)+3I_1(\Omega_z)+6I_1(\Omega_W)+I(\log\Omega_z+2\log\Omega_W)+
\nonumber\\
&+&\left[\frac{\varphi^2+I_0(\Omega_h)}{4}g^2-\Omega_W^2\right]J(\Omega_W)+
\frac{1}{2}
\left[\frac{\varphi^2+I_0(\Omega_h)}{4}(g^2+{g^\prime}^2)-\Omega_Z^2\right]J(\Omega_Z)+
\nonumber\\
&+&\left[\frac{9}{4}e^2I_0(0)+\frac{3}{8}g^2J(\Omega_W)
+\frac{3}{4}\frac{e^2g^2}{{g^\prime}^2}J(\Omega_z)\right]J(\Omega_W)
\label{GEPexpl}
\EA
where 
the function $J(X)$ is
\BE
J(X)=3I_0(X)+\frac{I}{X^2}
\label{J}
\EE
and the Euclidean integrals $I$, $I_0$, $I_1$ are defined according to
\BE
I=\int_\Lambda \frac{d_E^4k}{(2\pi)^4}
\label{I}
\EE
\BE
I_0(X)=\int_\Lambda \frac{d_E^4k}{(2\pi)^4}\frac{1}{k^2+X^2}
\label{I0}
\EE
\BE
I_1(X)=\frac{1}{2}\int_\Lambda \frac{d_E^4k}{(2\pi)^4}\log(k^2+X^2).
\label{I1}
\EE
Here the symbol $\int_\Lambda$ means that the integrals are regularized
by insertion of a cut-off $\Lambda$ so that $k<\Lambda$:
the Higgs sector is regarded as an effective model with a high energy scale
$\Lambda$ which plays the role of a further free parameter.

\section{The Gap Equations}

The variational parameters $\Omega_h$, $\Omega_W$ and $\Omega_Z$ must be determined
by requiring that for any value of $\varphi$ the GEP is at a minimum, thus the three parameters
are implicit functions of the average of the field $\rho$. Once the parameters
have been determined, the minimum point of  ${\cal V}_{GEP}$ as a function of $\varphi$
gives the vacuum expectation
value of the field $\rho$.
For any $\varphi$, the minimum of ${\cal V}_{GEP}$ is obtained by the constraints
\BE
\frac{\partial{\cal V}_{GEP}}{\partial\Omega_h^2}=
\frac{\partial{\cal V}_{GEP}}{\partial\Omega_W^2}=
\frac{\partial{\cal V}_{GEP}}{\partial\Omega_Z^2}=0.
\label{gap}
\EE
We find three coupled equations (gap equations) which define the implicit functions
$\Omega_h(\varphi)$, $\Omega_W(\varphi)$ and $\Omega_Z(\varphi)$.
Once the parameters have been set at their best value by solving the gap equations,
the broken-symmetry vacuum expectation value of the field $\rho$  takes
the value $\varphi_0$ which is obtained by the vanishing of the total derivative
\BE
\frac{d{\cal V}_{GEP}}{d\varphi}=\frac{\partial{\cal V}_{GEP}}{\partial\varphi}
+\sum_b\left(\frac{\partial{\cal V}_{GEP}}{\partial\Omega^2_b}\right)
\left(\frac{d \Omega^2_b}{d \varphi}\right)
\label{partials}
\EE
where the label $b$ runs over the bosons $W$, $Z$ and $h$.
If the gap equations are satisfied then
\BE
\frac{\partial{\cal V}_{GEP}}{\partial\Omega^2_b}=0
\EE
and the total derivative is equal to the partial derivative. Then $\varphi_0$
follows from the vanishing of the simple partial derivative
\BE
\left(
\frac{\partial{\cal V}_{GEP}}{\partial\varphi}
\right)_{\varphi=\varphi_0}=0.
\label{rho}
\EE
Eq.(\ref{gap}) and Eq.(\ref{rho}) are a set of four coupled equations that give the
phenomenological predictions of the model. Differentiating Eq.(\ref{GEPexpl})
the gap equations Eq.(\ref{gap}) can be written as
\BE
\Omega_h^2=m^2+\frac{\lambda}{2}\varphi^2+\frac{\lambda}{2} I_0(\Omega_h)+
\frac{g^2}{2}J(\Omega_W)+\frac{g^2+{g^\prime}^2}{4}J(\Omega_Z)
\label{GAPh}
\EE
\BE
\Omega_Z^2=(g^2+{g^\prime}^2) \frac{\varphi^2+I_0(\Omega_h)}{4}
+\frac{3e^2g^2}{2{g^\prime}^2}J(\Omega_W)
\label{GAPZ}
\EE
\BE
\Omega_W^2=g^2 \frac{\varphi^2+I_0(\Omega_h)}{4}
+\frac{3e^2g^2}{4{g^\prime}^2}J(\Omega_Z)
+\frac{9}{4} e^2 I_0(0)+\frac{3}{4}g^2J(\Omega_W).
\label{GAPW}
\EE
The vacuum expectation value of the field $\rho$ then follows from Eq.(\ref{rho}):
the partial derivative reads
\BE
\frac{\partial{\cal V}_{GEP}}{\partial\varphi}
=
\varphi
\left[m^2+\frac{\lambda}{6}\varphi^2+\frac{\lambda}{2} I_0(\Omega_h)+
\frac{g^2}{2}J(\Omega_W)+\frac{g^2+{g^\prime}^2}{4}J(\Omega_Z)\right]
\EE
and insertion of Eq.(\ref{GAPh}) yields
\BE
\frac{d{\cal V}_{GEP}}{d\varphi}=
\frac{\partial{\cal V}_{GEP}}{\partial\varphi}=
\varphi\left[\Omega_h^2-\frac{\lambda\varphi^2}{3}\right].
\label{rho2}
\EE
Then Eq.(\ref{rho}) has two solutions: the unbroken
symmetry stationary point $\varphi_0=0$ and the physical broken-symmetry vacuum
expectation value
\BE
\varphi_0^2=\frac{3}{\lambda}\Omega_h^2.
\label{rho3}
\EE
According, when $\varphi_0$ is set at its phenomenological value $v$,
the self-coupling constant $\lambda$ turns out to be proportional to the
square of the mass parameter $\Omega_h$, and a large $\Omega_h$
would not be compatible with perturbation theory. Conversely the present variational 
calculation still holds for any large coupling, allowing for a full discussion of
the Higgs sector. We notice that $\Omega_h$ is not the phenomenological mass 
$M_h$ of
the Higgs Boson which can be smaller than the variational parameter $\Omega_h$.
Here $\Omega_h$ may be regarded as the bare mass which appears in the zero-order
Lagrangian ${\cal L}_{GEP}(h)$ in Eq.(\ref{LGEPh}), and in principle it can be
very large.
The phenomenological mass of the Higgs boson arises from the curvature of the
GEP at the broken-symmetry minimum. 
Streactly speaking we should also check that the curvature is positive, otherwise
the solution of the coupled gap
equations would not refer to a minimum of the GEP.
At tree level, the perturbative result $M_h^2=\lambda \varphi_0^2/3$ would be
equivalent to Eq.(\ref{rho3}) only if $\Omega_h=M_h$. In fact we will see that
the bare mass $\Omega_h$ can be very large compared to
the Higgs boson mass
$M_h$  , and even a light
Higgs boson could be described by a strongly interacting Higgs sector with a
very large self-coupling $\lambda$\cite{siringo_light,siringo_var}. 

The curvature of the GEP follows from the second derivative: from Eq.(\ref{rho2})
we see that
\BE
\frac{d^2{\cal V}_{GEP}}{d\varphi^2}=
\left[\Omega_h^2-\frac{\lambda\varphi^2}{3}\right]
+2\varphi^2\left[\frac{d\Omega_h^2}{d\varphi^2}-\frac{\lambda}{3}\right]
\label{d2}
\EE
At the unbroken symmetry stationary point $\varphi_0=0$ the second term vanishes
\BE
M_0^2=\left(\frac{d^2{\cal V}_{GEP}}{d\varphi^2}\right)_{\varphi=0}
=\Omega_h^2
\label{M0}
\EE
and the physical mass is $M_0=\Omega_h$. Conversely in the phenomenological
broken-symmetry vacuum the first term vanishes and the physical mass $M_h$
is given by
\BE
M_h^2=\left(\frac{d^2{\cal V}_{GEP}}{d\varphi^2}\right)_{\varphi=\varphi_0}
=\frac{6\Omega_h^2}{\lambda}\left[
\left(\frac{d \Omega_h^2}{d\varphi^2}\right)_{\varphi=\varphi_0}
-\frac{\lambda}{3}\right].
\label{Mh}
\EE
The derivatives of the variational parameters $\Omega_b$ can be obtained 
by differentiating the coupled gap equations 
Eq.(\ref{GAPh}),(\ref{GAPZ}) and (\ref{GAPW}): we get the following
set of coupled linear
equations

\BE
\left[1-\frac{\lambda}{2}\frac{\partial I_0(\Omega_h)}{\partial\Omega_h^2}\right]
\left(\frac{d\Omega^2_h}{d\varphi^2}\right)
-\frac{1}{4}(g^2+{g^\prime}^2)\frac{\partial J(\Omega_Z)}{\partial \Omega_Z^2}
\left(\frac{d\Omega^2_Z}{d\varphi^2}\right)
-\frac{g^2}{2}\frac{\partial J(\Omega_W)}{\partial \Omega_W^2}
\left(\frac{d\Omega^2_W}{d\varphi^2}\right)=
\frac{\lambda}{2}
\EE

\BE
\frac{g^2}{4}\frac{\partial I_0(\Omega_h)}{\partial \Omega_h^2}
\left(\frac{d\Omega^2_h}{d\varphi^2}\right)
+\left(\frac{3e^2g^2}{4{g^\prime}^2}\right)
\frac{\partial J(\Omega_Z)}{\partial \Omega_Z^2}
\left(\frac{d\Omega^2_Z}{d\varphi^2}\right)
-\left[1+\frac{7 g^2}{4}\frac{\partial J(\Omega_W)}{\partial\Omega_W^2}\right]
\left(\frac{d\Omega^2_W}{d\varphi^2}\right)=
-\frac{g^2}{4}
\EE

\BE
\frac{1}{4}(g^2+{g^\prime}^2)\frac{\partial I_0(\Omega_h)}{\partial \Omega_h^2}
\left(\frac{d\Omega^2_h}{d\varphi^2}\right)
-\left(\frac{d\Omega^2_Z}{d\varphi^2}\right)
+\left(\frac{3e^2g^2}{2{g^\prime}^2}\right)
\frac{\partial J(\Omega_W)}{\partial \Omega_W^2}
\left(\frac{d\Omega^2_W}{d\varphi^2}\right)=
-\frac{1}{4}(g^2+{g^\prime}^2).
\EE

The solution is trivial, and insertion of $d\Omega_h^2/d\varphi^2$ in
Eq.(\ref{Mh}) yields the physical mass of the Higgs boson.

\section{Modified Variational Method and Phenomenology}

In order to explore the predictions of the model 
we have to face two major problems: we must 
find a consistent way to deal with the
diverging integrals, and we must renormalize the bare parameters
before any comparison with the phenomelogical observables can be made.

There are two very different approaches to the above problems.
While a consistent and satisfactory 
renormalization scheme will be described in the
next section, here we  discuss a simpler path which can be seen as
a modified low energy variational method.
This approach relies on the opinion that the Higgs sector
of the standard model is an effective model valid up to a
physical energy cut-off $\Lambda$. Thus the parameters in the
Lagrangian should be regarded as the physical effective values at that
energy scale. Besides, the energy scale $\Lambda$ could even be not too large
compared to the physical masses. According, the integrals can be
regularized by insertion of the cut-off, 
and the variational parameters $\Omega_b$ can be regarded
as phenomenological masses (with the eventual renormalization of $\Omega_h$
arising from Eq.(\ref{Mh}) above).
Actually the existence of strongly
diverging terms in the gap equations
(mainly the $J$ and $I$ integrals) makes it obvious that $\Lambda$
would give the scale of all the masses: in other words $\Lambda$ could not
be higher then $\approx100$ GeV. 
We could hardly find a physical meaning for such
a small scale, but rather $\Lambda$ should be regarded as a parameter that
cuts the high energy effects in the loop integrals: that allows the variational
method to describe the low energy physics without having to depend too much on the
high energy modes that usually spoil the predictive power of variational
calculations in field theory\cite{Feynman}. In that sense $\Lambda$ should not
be regarded as a truly physical parameter, but rather as an internal parameter
of the variational method. From a more formal point of view, that is equivalent
to split the GEP as
\BE
{{\cal V}_{GEP}}={\cal V}_{low}+{\cal V}_{high}
\EE
where ${\cal V}_{high}$ contains all the high energy contributions
to ${\cal V}_{GEP}$, i.e. all the contributions that arise from 
integrations over 
$k>\Lambda$ in Eqs.(\ref{J}),(\ref{I}),(\ref{I0}),(\ref{I1}).
We could define a modified variational method by taking the minimum
of the low energy part ${\cal V}_{low}$ only. The resulting gap equations
would be exactly the same as Eqs.(\ref{GAPh}),(\ref{GAPZ}) and (\ref{GAPW}) 
but with a cut-off $\Lambda$ in the diverging integrals.
Thus a small cut-off can be regarded as a special choice of an 
optimized variational method which enhances the effects of low energy modes.  

In this framework we may fix the masses
at their known phenomenological value, and look for a set of values
for the free parameters that satisfy the coupled gap equations. 
We can regard the masses $\Omega_W=80.403\pm0.029$ GeV and 
$\Omega_Z=91.1876\pm0.0021$ GeV as
experimentally known physical masses\cite{Data}. From the Fermi constant 
$G_F=1.16637\cdot 10^{-5}$ GeV$^{-2}$ the phenomenological 
weak coupling $g$ of the effective lagrangian follows according to
\BE
\frac{g^2}{\Omega_W^2}=4\sqrt{2} G_F
\EE
which yields $g=0.6531$. The fine-structure constant at the weak
scale\cite{weinberg2} reads
$e^2/4\pi=1/128.87$ and determines the weak coupling $g^\prime=0.3555$ by
Eq.(\ref{charge}). With these phenomenological inputs, we are left with
an unknown mass, namely the Higgs boson mass parameter $\Omega_h$, and the
three free parameters $\lambda$, $m^2$, $\Lambda$ that characterize the Higgs sector.

A first test of the model arises from a comparison of the known phenomenological
energies $v_W$ and $v_Z$, that we define as
\BE
\Omega_Z=\frac{1}{2} v_Z\sqrt{g^2+{g^\prime}^2} 
\EE
\BE
\Omega_W=\frac{1}{2}g v_W.
\EE
The phenomenological data yield $v_W=1/\sqrt{(\sqrt{2} G_F)}=246.221$ GeV,
$v_Z=245.264$ GeV and 
$(v_W-v_Z)/v_W\approx (v_W^2-v_Z^2)/(2v_W^2)=3.88\cdot 10^{-3}\pm 0.4\cdot 10^{-3}$.

In the standard model, at tree-level
both those energies are equal to the vacuum expectation value of the scalar
field $\rho$. The small phenomenological difference arises from higher
order corrections. In the present variational approximation the energies 
would differ
according to the gap equations Eq.(\ref{GAPZ}) and Eq.(\ref{GAPW}), 
and nor of them would be exactly
equal to $\varphi_0$:
\BE
v_Z^2=\varphi_0^2+I_0(\Omega_h)+6\cos^4 \theta_W 
J(\Omega_W)
\label{vz}
\EE
\BE
v_W^2=\varphi_0^2+I_0(\Omega_h)+3\cos^2 \theta_W
J(\Omega_Z)
+9\sin^2 \theta_W I_0(0)+3J(\Omega_W)
\label{vW}
\EE
where as usual we take $\cos^2\theta_W=e^2/g^2$ and
$\sin^2\theta_W=e^2/{g^\prime}^2$.
The difference arises from one-loop contributions, and does not depend on
the mass of the Higgs boson or on other parameters of the Higgs sector:
\BE
v_W^2-v_Z^2=9\sin^2\theta_W I_0(0)
+3\cos^2\theta_W J(\Omega_Z)
+3(1-2\cos^4\theta_W)J(\Omega_W)
\label{deltav2}
\EE
Of course this difference is sensitive to the magnitude of the cut-off $\Lambda$,
and can be regarded as a phenomenological constraint on the cut-off. In the
limit $\Lambda\to 0$ all the integrals in Eq.(\ref{deltav2}) vanish and we are left
with the tree-level approximation $v_Z=v_W=\varphi_0$. In the limit $\Lambda\to\infty$
all the integrals diverge and the calculation has no pratical meaning.

\begin{figure}[ht]
\includegraphics[height=10cm, width=10cm, angle=-90]{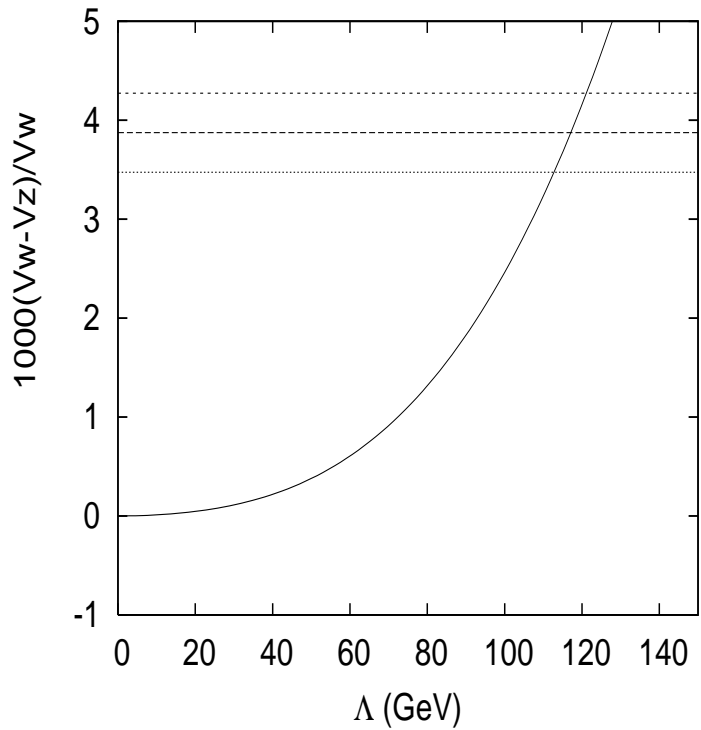}
\caption{\label{Fig1}
The relative difference $(v_W-v_Z)/v_W\approx(v_W^2-v_Z^2)/(2v_W^2)$ is reported
as a function of the cut-off $\Lambda$ according to Eq.(\ref{deltav2}).
The average phenomenological value is reported as a dashed line, 
and it is expected to be confined between the dotted lines.}
\end{figure}

Thus the difference
in Eq.(\ref{deltav2}) can be regarded as a 
measure of the contribution of high energy modes, and the phenomenological
requirement of keeping this difference at a small reasonable value is a strong
constraint.
In other words we can regard $\Lambda$ as a free parameter and use Eq.(\ref{deltav2})
in order to establish its value by comparison 
with the phenomenological value of the difference $v_Z^2-v_W^2$.
In condensed matter\cite{marotta,camarda} this freedom has allowed a good fit of the
experimental data, since the cut-off acts as a regulator of quantum fluctuations
that can be scaled at the correct phenomenological value. However we do not expect
that the parameter $\Lambda$ might retain any physical role besides being a fit
parameter, and its actual value cannot be taken too seriously as a physical
energy scale. In fact, as expected, 
it turns out to be quite small, thus indicating that
quantum fluctuations do not play a major role. 

The relative difference is evaluated by Eq.(\ref{deltav2}) and
reported in Fig.1 as a function of the cut-off $\Lambda$.
It is an increasing function of $\Lambda$ and
it reaches the phenomenological value for $\Lambda\approx 115$ GeV. For larger
values of $\Lambda$ the quantum fluctuations add a large increasing contribution
which eventually diverges. Then it is quite clear that,
in this framework, the variational method does not introduce any 
dramatic change with respect to the tree-level description, since
the role of quantum fluctuations is
strongly suppressed by the small value of $\Lambda$.
This result does not seem to be very satisfactory as according to Eq.(\ref{Mh})
any renormalization of the Higgs Boson mass would be negligible and
strongly suppressed by the small cut-off: in order to keep the difference 
between $v_W$ and $v_Z$ small, any interesting effect would be suppressed and
the method would be equivalent to tree-level perturbation theory.

As the problem arises from the existence of strongly diverging integrals in finite
phenomenological observables, and since those dievergences are known to be 
spurious (i.e. they cancel exactly at one-loop\cite{Appelquist,Kummer}), 
we conclude that a satisfactory comparison with the phenomenological data
requires a deeper and consistent renormalization of the physical observables.

\section{Renormalization and Phenomenological Predictions}

In order to overcome the shortcomings of the previous section, we
develop a consistent renormalization scheme which does not spoil the variational
nature of the method in the Higgs sector, while retaining the structure of a
standard perturbative calculation for the weak interactions. 
The 
variational results are regarded as the starting point of
a standard perturbative renormalization process. This view had been suggested
by Cea and Tedesco\cite{Cea} who had also shown that the gaussian variational
trial functionals can be regarded as an optimized variational basis.
Higher order corrections can be derived by standard perturbative techniques in
the variational basis, still retaining the variational nature of the calculation.
In the zero-order
Lagrangian ${\cal L}_{GEP}$, Eq.(\ref{LGEPTOT}), 
the variational evaluation of the mass parameters $\Omega_b$  
ensures that the residual interaction
is minimal, and the so optimized perturbative theory should work better.
We do not expect any important improvement in the weak sector 
where the couplings
are very small and 
the standard perturbative theory already works very well. 
However the method gives new insights on the Higgs sector,
and on its non-perturbative range of strong coupling.

We take the view that all the variational parameters $\Omega_b$ must be regarded
as bare parameters which define the bare optimized zero-order Lagrangian
${\cal L}_{GEP}$. Thus no comparison can be done with experimental masses before
renormalization. The cut-off $\Lambda$ is regarded as a large energy scale of
the effective model, and the bare masses $\Omega_b$ are derived by solution
of the gap equations Eq.(\ref{GAPh}),(\ref{GAPZ}),(\ref{GAPW}).
These bare masses would be very large, typically of order $\Lambda$, and must
be renormalized. For the weak sector we can use 
the standard one-loop renormalization of masses, while for the Higgs sector a
consistent non-perturbative renormalization is required, and it is provided
by the curvature of the GEP Eq.(\ref{Mh}) which 
contains non-perturbative contributions to the mass shift. Actually in the
simple scalar theory Eq.(\ref{Mh}) can be shown to be equivalent to
the sum of bubble diagrams to all orders\cite{siringo_bubble,siringo_var}.
Thus the shape of the GEP contains non-perturbative information on the 
renormalization of the bare parameters. In the Higgs sector it is quite 
important that we rely on a non-perturbative renormalization method as 
we would like to discuss some untrivial features of the strong coupling regime.

In the weak sector all divergences are known to cancel at one-loop even in the
unitarity gauge\cite{Appelquist,Kummer} and the resulting perturbative corrections 
have been reported to be very small, as it should be for any perturbative 
correction arising from weak couplings. We can recover the same results from the
optimized zero-order Lagrangian ${\cal L}_{GEP}$ with bare masses $\Omega_b$ that are
solution of the gap equations and with $\varphi$ set at the minimum of the GEP
$\varphi=\varphi_0=v$ which we expect to be the phenomenological value. 
The one-particle irreducible one-loop self-energy
reads
\BE
\Sigma_b^{(1L)}=\Omega^2_b-M_b^2+\Sigma_b^{(1)}+\Sigma_b^{(2)}
\label{self}
\EE
where the index $b$ runs over the two bosons $W$ and $Z$, while
$\Sigma^{(1)}$ and $\Sigma^{(2)}$ are the first and second order
contributions to the one-loop self-energy, and the masses $M_b$ are the standard
tree level masses as reported in Eqs.(\ref{MW}),(\ref{MZ}). It can be easily
shown that the gap equations Eqs.(\ref{GAPZ}),(\ref{GAPW}) can be written in terms
of the first order self energy as
\BE
\Omega_b^2=M_b^2-\Sigma_b^{(1)}
\label{GAPS}
\EE
and insertion in Eq.(\ref{self}) shows that
$\Sigma_b^{(1L)}=\Sigma_b^{(2)}$ which is a general property
of the GEP.
At one-loop the renormalized mass is
\BE
(\Omega^2_b)_R=\Omega_b^2-\Sigma_b^{(1L)}
\EE
and again, insertion of the gap equation Eq.(\ref{GAPS})
yields
\BE
(\Omega^2_b)_R=M_b^2-\Sigma_b^{(1)}-\Sigma_b^{(2)}.
\EE
Formally this is exactly the same result which we would obtain by 
one-loop renormalization of the standard model lagrangian with bare
masses $M^2_b$ (apart from the choice of the free propagator in the self energy, 
which in the present calculation contains the bare masses $\Omega_b$).
The sum of all one-loop terms contributing to $\Sigma_b^{(1)}+\Sigma_b^{(2)}$ 
is known to be finite and very small\cite{Appelquist,Kummer} compared to $M_b^2$.
Thus we get the standard one-loop result $(\Omega^2_b)_R\approx M_b^2$
up to small perturbative corrections.
We can say that the one-loop renormalization of the bare variational masses
allows us to recover the experimental phenomenology for the gauge bosons.
The result would be trivial were it not for the Higgs sector where 
the coupling cannot be assumed to be small and where the above 
perturbative renormalization would not be reliable.
In the Higgs sector the variational mass parameter $\Omega_h$ depends on
the self-coupling $\lambda$ and on the vacuum expectation value $\varphi_0=v$ through 
the minimum condition Eq.(\ref{rho3}) which gives to $\Omega_h$ a clear physical
phenomenological meaning: $\Omega_h$ sets the scale of the self-coupling
$\lambda$ which reads 
\BE
\lambda=\frac{3\Omega_h^2}{v^2}
\label{coupl}
\EE
Here we do not have any problem at insuring that    
$\Omega_h$, 
the solution of the gap equation Eq.(\ref{GAPh}), 
takes a finite
phenomenological value: in fact the existence of the free mass parameter $m^2$
makes sure that the solution of the gap equation Eq.(\ref{GAPh}) can be any
number we like. Thus we fix $m^2$ in order to satisfy the minimum condition
Eq.(\ref{coupl}) and take $\Omega_h$ as a free parameter which gives
the strength of the self-coupling $\lambda$. We do not need to deal 
with infinities, but we must address the problem of mass renormalization anyway,
as the residual interaction shifts the physical Higgs mass that cannot be
taken to be equal to the variational mass parameter $\Omega_h$.
In fact we have seen that the shape of the GEP contains non-perturbative
effects which can be shown to be the sum of bubble diagrams to all 
orders\cite{siringo_bubble,siringo_var}. 
A non-trivial mass renormalization comes from the curvature of the GEP
which allows us to take the physical mass of the Higgs boson $M_h$ 
according to Eq.(\ref{Mh}). At this stage there is no reason why
the cut-off $\Lambda$ should be small, since all the phenomenological observables
are finite any way. We assume that $\Lambda$ is some very large energy scale 
and examine the behaviour of the physical mass $M_h$ as a function of
the self-coupling $\lambda$.

\begin{figure}[ht]
\includegraphics[height=10cm, width=10cm, angle=-90]{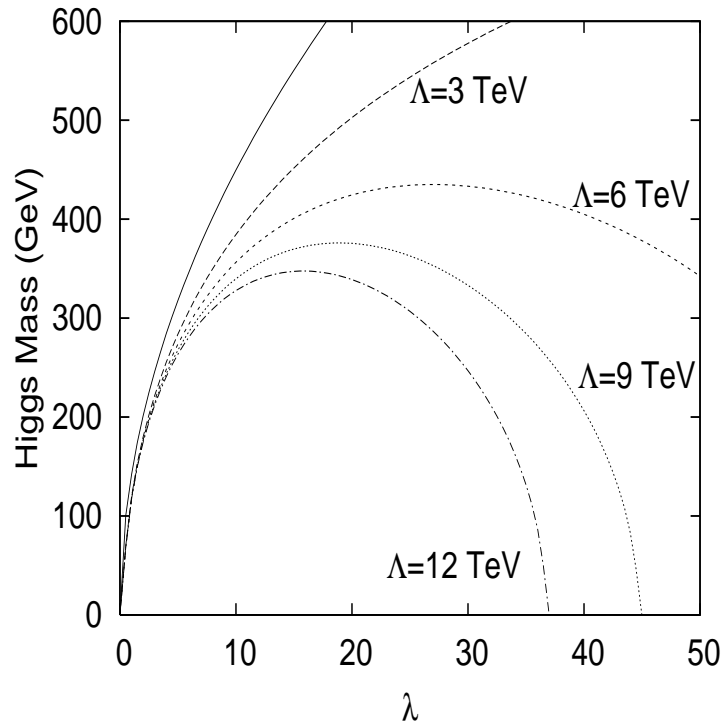}
\caption{\label{Fig2}
The Higgs mass $M_h$ according to Eq.(\ref{Mh}) as a function of the self-coupling
parameter $\lambda$ and for several choices of the cut-off $\Lambda$ ranging
from 3 TeV to 12 TeV (broken lines). The solid line is the tree-level result
$M_h=\Omega_h$ ($\Lambda=0$).}
\end{figure}

\begin{figure}[ht]
\includegraphics[height=10cm, width=10cm, angle=-90]{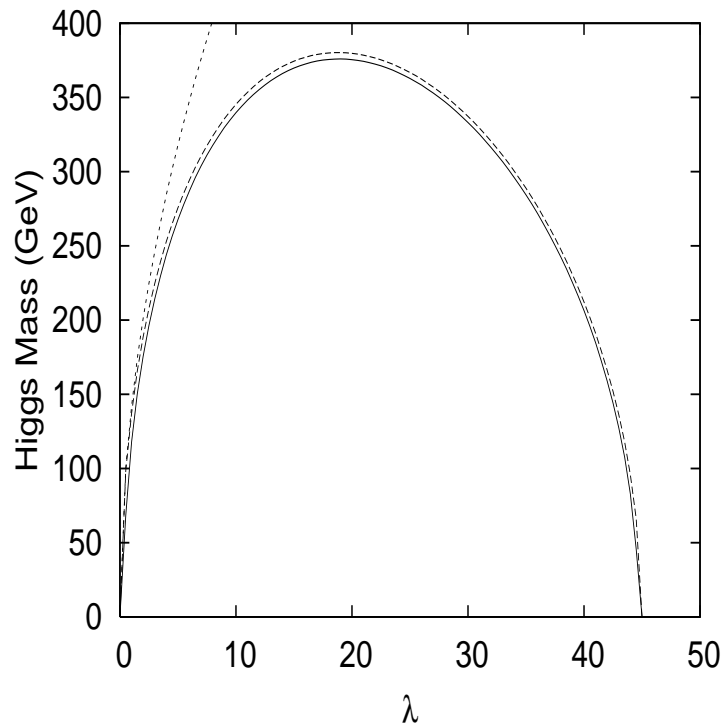}
\caption{\label{Fig3}
The Higgs mass $M_h$ according to Eq.(\ref{Mh}) as a function of the self-coupling
parameter $\lambda$ for $\Lambda=9$ TeV (solid line).
The dashed line is the simple scalar theory result\cite{siringo_light,siringo_bubble}. 
For comparison
the tree-level result 
$M_h=\Omega_h$ ($\Lambda=0$) is reported as a dotted line.}
\end{figure}

As shown in Fig.2 the physical mass $M_h$ is not a monotonous increasing function
of the coupling, but it reaches a maximum and then decreases. $M^2_h$ eventually 
becomes negative at some large coupling, indicating that the broken-symmetry solution
becomes unstable. We get an upper bound for the coupling and, before reaching it, 
a low mass non-perturbative strong coupling range. In this scenario a light Higgs
can be found for a small coupling (perturbative light Higgs) but also 
for a large coupling (non-perturbative light Higgs). A very strong self-coupling
reduces the mass: this effect cannot be predicted by any perturbative calculation.
Moreover the reduction of mass increseas with the increasing of the 
cut-off $\Lambda$ and eventually
an infinite cut-off would make the Higgs boson mass vanishing: the broken symmetry
vacuum would become unstable for any coupling as the upper bound of $\lambda$ would
go to zero. That is in agreement with the well known triviality of the scalar
theory which requires the existence of a large but finite cut-off. 
In Fig.2 the tree-level approximation $M_h=\Omega_h$ is also reported for comparison:
it is equivalent to the variational calculation for a very small cut-off $\Lambda$ 
as discussed in the previous section. We can see that in the perturbative regime
of small $\lambda$ the Higgs mass is almost insensitive to the size of the cut-off,
and the perturbative predictions agree with the variational result: the mass 
increases as the square root of the self-coupling $\lambda$. Conversely, in
the strong coupling regime the mass of the Higgs boson depends on the size of the
cut-off and becomes very small compared to the perturbative
prediction which cannot be trusted any more.
For instance at $\Lambda=12$ TeV, a relatively light Higgs boson with 
$M_h\approx200$ GeV 
is predicted for $\lambda\approx 2.5$ (perturbative weak-coupling range) but also for
$\lambda\approx40$ (non-perturbative strong-coupling range). 
The mass is the same in both cases
but we expect a different behaviour for the scattering amplitudes in the 
strong-coupling range\cite{siringo_var}. 

The prediction of a light Higgs boson in the strong coupling regime had been
discussed in simplified models which neglected the gauge 
interactions\cite{siringo_var,siringo_light} and in the Abelian gauge interacting
U(1) theory\cite{ibanez}. Here we confirm the same trend in
the framework of the full SU(2)$\times$U(1) gauge theory. 

In Fig.3 the prediction
of the GEP for a simple scalar theory\cite{siringo_light,siringo_bubble} 
is reported for comparison. As expected,
the effect of gauge interactions is very small and can be neglected for a qualitative
discussion of the Higgs sector.

\begin{figure}[ht]
\includegraphics[height=10cm, width=10cm, angle=-90]{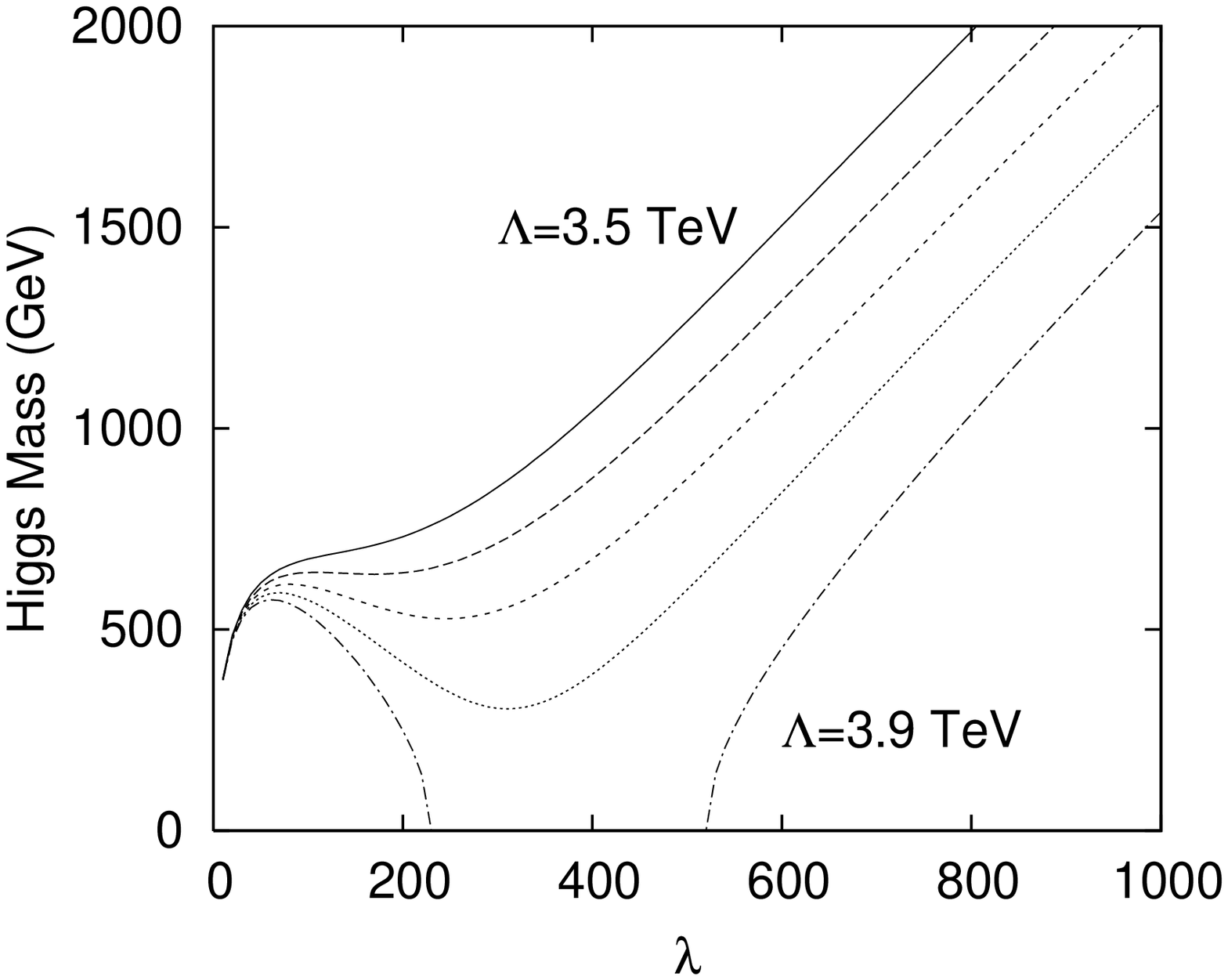}
\caption{\label{Fig4}
The Higgs mass $M_h$ according to Eq.(\ref{Mh}) as a function of the self-coupling
parameter $\lambda$ for $\Lambda$=3.5, 3.6, 3.7, 3.8 and 3.9 TeV.
}
\end{figure}

Quite interesting the physics of the Higgs sector changes according to the choice
of the cut-off $\Lambda$ with a cross over point at $\Lambda\approx 3.7$ TeV
separating the "small cut-off" scenario from the "large cut-off" one.
As shown in Fig.4 the Higgs boson mass is an increasing function of the self-coupling
$\lambda$ for $\Lambda<3.5$ TeV. At $\Lambda\approx 3.7$ TeV a minimum appears and
deepens until $\Lambda\approx 3.9$ Tev where the minimum of $M_h^2$ becomes negative
and a gap of prohibited couplings opens. 
We notice that the existence of a similar cross-over had been 
observed at $\Lambda\approx 3$ Tev 
in a quite different variational calculation for the scalar theory\cite{siringo_var}.
Thus it seems to be a genuine feature of the standard model which cannot be shown by any
perturbative approximation.

In the "large cut-off" scenario, say for
$\Lambda> 3.9$ TeV, there is a gap in the allowed range of $\lambda$, and this
gap increases with the increasing of the cut-off. In this scenario a light Higgs boson,
with a very small mass, can be compatible with three different couplings: for instance
at $\Lambda\approx 3.9$ TeV (as shown in Fig.4) these are $\lambda_1\to 0$ 
(perturbative solution),
$\lambda_2\approx 230$ and $\lambda_3\approx 520$.
Of course any other choice for $\Lambda$ would provide a different set of couplings
for the same Higgs mass. For a very large cut-off $\Lambda$ the bigger coupling
$\lambda_3$ becomes very large and probably has no physical relevance, while the
intermediate coupling $\lambda_2$ becomes smaller and would describe a strongly
coupled Higgs sector. The general behaviour is shown in Fig.2 where at 
$\Lambda=12$ TeV the intermediate coupling reduces to $\lambda_2\approx 37$.
Thus for a large enough cut-off $\Lambda$ we can predict the possible existence
of a light Higgs boson with a large but reasonable self-coupling $\lambda_2$.

The possible existence of a light Higgs boson with a very strong self-interaction
seems to be a non-perturbative feature of the standard 
model\cite{siringo_var,siringo_light, ibanez}. Thus the eventual experimental 
finding of a light Higgs mass $M_h\approx 200$ TeV would not rule out a strongly
interacting Higgs sector. However we expect that a strongly interacting light
Higgs boson should show a different behaviour when compared with the perturbative
predictions: scattering amplitudes should be different and should tell us about
the real strength of the self-coupling\cite{note}.


\begin{thebibliography} {99}
\bibitem{note} Of course any direct comparison would probably require a consistent 
non-perturbative renormalization of the coupling $\lambda$ along the lines discussed
in Ref.\cite{siringo_bubble}
\end{thebibliography}
\end{document}